\documentclass[preprint,12pt]{revtex4-2}
\usepackage{amsmath,amssymb}
\usepackage{graphicx}
\usepackage{dcolumn}
\usepackage{bm}
\usepackage{siunitx}
\usepackage{booktabs}
\usepackage{mhchem}
\usepackage{float}
\usepackage{hyperref}
\hypersetup{
    colorlinks=true,
    linkcolor=blue,
    citecolor=blue,
    urlcolor=blue
}

\begin{document}

\title{\bf Resolution of the Two-Dimensional Ferromagnetic Spin-3/2 Ising Model via Cluster Growth}

\author{J. Roberto Viana$^{a}$}
\author{Octavio D. Rodriguez Salmon$^{a}$}
\author{Minos A. Neto$^{a}$}
\author{Griffith Mendonça$^{a}$}
\author{F. Dinóla Neto$^{a}$}
\affiliation{$^{a}$ Departamento de F\'{\i}sica, Universidade Federal do Amazonas, 3000, Japiim, 69077-000, Manaus-AM, Brazil
}

\begin{abstract}
We propose a computational methodology based on a hierarchical cluster growth process to solve spin-3/2 Ising models efficiently. The method circumvents the exponential complexity (\(4^{N}\)) of the canonical ensemble partition function by iteratively constructing finite magnetic clusters of size \(N_g\), where the effective spin state of a site in generation \(g+1\) is determined by the local magnetization of a cluster from generation \(g\). This approach, which shares conceptual ground with effective field theories, allows the study of systems of effectively very large size \(N = N_0 (N_g)^{g}\). We apply the formalism to the ferromagnetic spin-3/2 Ising model on a honeycomb lattice, modeling the monolayer CrI$_3$,  a prototypical two-dimensional Ising magnet. The model, calibrated using the experimental transition temperature (\(T_{c} \simeq 45\) K), successfully reproduces key experimental features: the temperature dependence of the magnetization \(m(T)\), including its inflection point, and the broadened peak in the specific heat \(c_v(T)\). We also compute the entropy \(s(T)\), finding a finite residual value at low temperatures consistent with the system's double degeneracy. Our results demonstrate that this hierarchical cluster method provides a quantitatively accurate and computationally efficient framework for studying complex magnetic systems.


\end{abstract}
\maketitle

\section{Introduction}

The development of theoretical models for the study of magnetic systems has played a central role in the advancement of Condensed Matter Physics, providing fundamental insights into the microscopic origin of magnetism in materials.
These theoretical developments have not only deepened our understanding of magnetic phenomena, but have also enabled significant technological progress through the design and optimization of magnetic materials.

Among the theoretical approaches used to describe magnetic systems, spin models occupy a prominent position. In particular, the Ising and Heisenberg models have served as paradigmatic frameworks for investigating ferromagnetic
ordering, phase transitions, and critical phenomena. These models allow for a systematic analysis of thermodynamic and magnetic properties, facilitating direct qualitative and quantitative comparisons between theoretical predictions
and experimental observations.

Within this context, the Ising model represents a cornerstone in the theory of ferromagnetism in discrete systems. A landmark result was obtained by Onsager, who derived the exact solution for the critical temperature of the spin-$1/2$ Ising model on a two-dimensional lattice~\cite{1}. This work established a
reference point for subsequent analytical and numerical investigations of magnetic criticality.

Over the years, a wide range of theoretical techniques has been developed to study magnetic systems, including Renormalization Group approaches~\cite{2,3}, cluster-based mean-field theories~\cite{4,5}, spin-wave theory~\cite{6}, Monte
Carlo simulations~\cite{7,8,9}, and numerical methods for quantum spin systems~\cite{10,11}. These methods have provided a comprehensive understanding of thermodynamic behavior, critical exponents, and excitation spectra in
various magnetic models.

In parallel, extensive experimental efforts have been devoted to the study of ferromagnetic materials in different physical realizations. Experimental investigations have addressed phase transition temperatures, hysteresis effects, magnetic domain formation, magnetocrystalline anisotropy, strong spin-electron coupling, weak ferromagnetism, and thin-film magnetism
\cite{12,13,14,15,16,17,18,19,20}. These studies have highlighted the importance of dimensionality, symmetry, and interaction range in determining magnetic behavior.

More recently, renewed interest in low-dimensional magnetism has been driven by the experimental discovery of intrinsic ferromagnetism in atomically thin materials. Coldea \textit{et al.}~\cite{21} investigated a quasi-one-dimensional
Ising system, while Huang \textit{et al.}~\cite{22} and Gong \textit{et al.}~\cite{23} independently reported ferromagnetic order in monolayers of van der Waals materials, such as CrI$_3$ and Cr$_2$Ge$_2$Te$_6$. Subsequent studies have explored a variety of two-dimensional magnetic systems, including
kagome lattices with Heisenberg-like behavior and magnonic
excitations~\cite{24}.

From a theoretical perspective, modeling magnetic systems with higher spin values and realistic length scales remains a challenging task. For instance, a spin-$3/2$ system possesses $4^{N}$ accessible states, which renders an exact treatment computationally prohibitive for large system sizes. While numerical methods and cluster-based approaches can capture local correlations, extending these techniques to mesoscopic or macroscopic scales often requires additional approximations.

Motivated by these challenges, in this work we propose a multiscale theoretical approach for the study of magnetic systems based on the generation of effective clusters. The method combines ideas from Effective Field Theory and
Renormalization Group concepts, allowing the system to evolve through successive generations of clusters. At each scale, the thermodynamic behavior is obtained by solving a finite-size canonical ensemble problem, while effective magnetizations and coupling constants are recursively renormalized to encode the influence of larger length scales.

As a concrete application, we apply the proposed framework to a ferromagnetic Ising-type system with spin $3/2$ defined on a two-dimensional lattice with honeycomb topology. The theoretical results are compared with experimental data
for the magnetic material CrI$_3$ (chromium triiodide), which was one of the first materials shown to exhibit intrinsic ferromagnetism in a single atomic layer~\cite{22,23}. This discovery was particularly remarkable in light of the
Mermin-Wagner theorem, which forbids spontaneous long-range magnetic order in two-dimensional isotropic systems at finite temperature. In monolayer CrI$_3$, however, strong magnetocrystalline anisotropy leads to an effective Ising-like
behavior ($\mathbb{Z}_2$ symmetry), thereby allowing long-range magnetic order to exist in two dimensions.

Experimentally, CrI$_3$ exhibits Curie temperatures of approximately $T_{\mathrm{C}} = 45$~K in the monolayer limit and $T_{\mathrm{C}} = 61$~K in the bulk~\cite{22,23}. Its magnetic behavior depends sensitively on thickness: while
intralayer interactions are ferromagnetic, interlayer couplings can be antiferromagnetic, leading to a rich variety of magnetic phases in multilayer structures. Beyond its fundamental interest, CrI$_3$ is also a promising platform for two-dimensional spintronic applications, including ultrathin
magnetic devices, memories, and van der Waals heterostructures~\cite{27,28}.

The proposed framework is conceptually inspired by
renormalization-group ideas in the sense that it employs a hierarchical coarse-graining across scales. However, it differs fundamentally from standard real-space RG schemes. In particular, no degrees of freedom are integrated out,
no explicit length rescaling is performed, and no flow in the space of Hamiltonians is constructed. Instead, the present approach operates directly at the level of thermodynamic observables through a phenomenological scale-dependent coupling, calibrated using experimental input. For this
reason, the method should be regarded as RG-inspired rather than a renormalization-group procedure in the conventional sense.
 The remainder of this article is organized as follows. In Section II, we present the model and the theoretical framework, introducing the hierarchical cluster-growth formalism and the effective Hamiltonians that describe the successive generations of magnetic clusters. In Section III, we discuss how the effective system size is defined within the proposed approach and how the number of particles is related to the cluster generations and domain volume. In Section IV, we establish the recurrence relations between evolution scales, including an ansatz for the magnetization and for the effective magnetic coupling. In Section V, we analyze the robustness of the recurrence ansatz by considering generalized forms of the scale function and the corresponding physical constraints. In Section VI, we present and discuss the numerical results, including magnetization, internal energy, specific heat, entropy, and critical-exponent estimates, with application to the CrI$_3$ compound. Finally, in Section VII, we summarize the main conclusions and perspectives of the proposed method.

\section{Model and Theoretical Framework}

The theoretical framework proposed in this work is based on a dynamical evolution process defined in terms of successive generations of clusters, denoted by $g$. The procedure starts from a finite microscopic cluster of size $N_{0}$ composed of spin variables $S_{j}$, which defines the zeroth generation
($g=0$) of the system. This initial cluster represents the smallest scale at which the magnetic degrees of freedom are explicitly treated.

\begin{figure}[H]
    \centering
    \includegraphics[width=0.6\linewidth]{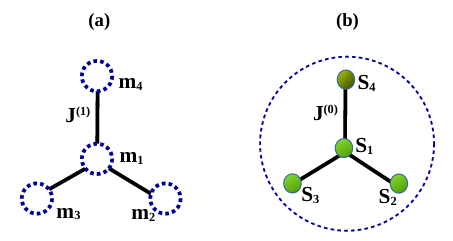}
    \caption{In (a) the first-generation cluster $g=1$ is illustrated, where each $m_j$ corresponds to a cluster mean. In (b) is the zero-generation cluster is depicted. }
    \label{fig:fig1}
\end{figure}

The dynamical growth of the system is implemented by constructing higher generations of clusters, where the degrees of freedom are progressively coarse-grained. Fig. \ref{fig:fig1} schematically illustrates the construction of a
first-generation cluster ($g=1$). In this representation, each site of the new cluster corresponds to a local magnetization variable $m_{j}$ obtained as an average over the spin degrees of freedom of the previous generation. In this
case, the number of sites is preserved, i.e., $N_{1}=N_{0}$, and the cluster topology remains unchanged.

For general generations $g \geq 1$, the system is described in terms of local magnetizations $m_{j}^{(g)}$, which encode the collective magnetic behavior of the cluster belonging to the previous generation $g-1$. In this way, the
degrees of freedom evolve from microscopic spin variables to effective mesoscopic quantities, allowing the description of increasingly larger length scales while keeping the computational complexity under control.

During the evolution of the cluster generations, the effective number of particles contributing to the thermodynamic and magnetic properties of the system is given by
\begin{equation}
N = N_{0} \left( N_{g} \right)^{g}.
\label{eq:eq0}
\end{equation}
The growth process is terminated at a maximum generation $g_{\max}$, which is determined by the characteristic volume of the magnetic domain under study. Each hypothetical magnetic domain is therefore associated with a cluster of generation $g_{\max}$.

The motivation for this modeling strategy is rooted in the principles of Effective Field Theory (EFT), where the objective is to capture the essential macroscopic behavior of real magnetic materials without requiring a complete microscopic description. In the vicinity of criticality, the correlation
length $\xi$ becomes much larger than the lattice constant, rendering microscopic details irrelevant and justifying a coarse-grained description in terms of effective variables.

Within this framework, the relevant physical ingredients are the effective interactions between sites, the local order parameter (magnetization), the symmetry of the system (Ising, XY, or Heisenberg), the dimensionality, and the interaction range. The proposed cluster-generation approach naturally
incorporates these elements and provides a systematic route to connect microscopic spin models to mesoscopic and macroscopic magnetic behavior.

Particularly, we consider a ferromagnetic Ising-type system with spin $3/2$ and nearest-neighbor interactions. The microscopic Hamiltonian corresponding to the zeroth generation ($g=0$) is given by
\begin{equation}
\mathcal{H}^{(0)} = -J_{1}^{(0)} \sum_{\langle i,j \rangle} S_{i}^{z} S_{j}^{z},
\label{eq:eq1}
\end{equation}

where $J_{1}^{(0)}$ denotes the ferromagnetic nearest-neighbor coupling constant and $S_{i}^{z} \in \{\pm 3/2, \pm 1/2\}$. Since the present study focuses on static properties, we set $\hbar \equiv 1$.

For generations $g \geq 1$, the microscopic spin variables are replaced by local magnetizations, and the effective Hamiltonian assumes the form 
\begin{equation}
\mathcal{H}^{(g)} = -J_{1}^{(g)} \sum_{\langle i,j \rangle}
m_{i}^{(g)} m_{j}^{(g)}.
\label{eq:eq2}
\end{equation}
Introducing the inverse temperature $\beta = 1/(k_{B}T)$, the dimensionless Hamiltonian can be written as
\begin{equation}
-\beta \mathcal{H}^{(g)} = K^{(g)} \sum_{\langle i,j \rangle}
m_{i}^{(g)} m_{j}^{(g)},
\label{eq:eq3}
\end{equation}
with
\begin{equation}
K^{(g)} = \frac{J_{1}^{(g)}}{k_{B} T}.
\label{eq:eq4}
\end{equation}
For convenience, we also define the temperature-independent parameter
\begin{equation}
\gamma^{(g)} \equiv \frac{J^{(g)}}{k_{B}},
\label{eq:eq5}
\end{equation}
so that $\gamma^{(0)}$ sets the initial energy scale of the magnetic coupling.

The local magnetization at each site and generation is defined through the
recursive relation
\begin{equation}
m_{j}^{(g)} =
F\!\left(m_{j}^{(g-1)}, J^{(g-1)}, T\right) s_{j},
\label{eq:eq6}
\end{equation}
where $s_{j} \in \{\pm 3/2, \pm 1/2\}$ and the function $F$ encodes the memory of the previous generation.

For the first generation the magnetization $m^{(0)}$ is obtained from the initial finite cluster as
\begin{equation}
m^{(0)} =
\pm \frac{1}{N_{0}}
\frac{1}{Z_{0}}
\sum_{\{S_{j}\}}
\left|
\sum_{j} S_{j}
\right|
\exp\!\left( -\beta \mathcal{H}^{(0)} \right),
\label{eq:eq7}
\end{equation}
where $Z_{0}$ is the canonical partition function of the zeroth-generation cluster.

For general generations $g \geq 1$, the magnetization is given by
\begin{equation}
m^{(g)} =
\pm \frac{1}{N_{g}}
\frac{1}{Z_{g}}
\sum_{\{m_{j}^{(g)}\}}
\left|
\sum_{j} m_{j}^{(g)}
\right|
\exp\!\left( -\beta \mathcal{H}^{(g)} \right),
\label{eq:eq8}
\end{equation}
with the corresponding canonical partition function
\begin{equation}
Z_{g} =
\sum_{\{m_{j}^{(g)}\}}
\exp\!\left( -\beta \mathcal{H}^{(g)} \right).
\label{eq:eq9}
\end{equation}

\section{System Size of the Particle Ensemble}

Within the proposed formalism, the effective number of particles involved in the magnetic domain represented by generation $g$ is given by
\begin{equation}
N = N_{0} \left( N_{g} \right)^{g}.
\label{eq:eq10}
\end{equation}

This formulation allows the thermodynamic properties of a system containing $N$ particles to be accessed without explicitly treating the full Hilbert space of dimension $4^{N}$. Instead, only $4^{N_{g}}$ configurations are
required at each stage of the growth process, which constitutes one of the main advantages of the present approach.

Real magnetic materials are generally inhomogeneous and consist of multiple magnetic domains with different magnetization orientations. In this first application of the proposed method, we restrict our analysis to a hypothetical system composed of a single magnetic domain with a well-defined volume.

The number of particles $N$ associated with a material containing $n$ mols is given by Avogadro's relation
\begin{equation}
N = n \times N_{A},
\label{eq:eq11}
\end{equation}
where $N_{A}$ is the Avogadro's number ($6.023 \times 10^{23}$). Equating this expression with the effective particle number of the model, the maximum generation required to represent the domain is obtained from
\begin{equation}
N_{0} \left( N_{g} \right)^{g_{\max}}
= n \times N_{A},
\label{eq:eq12}
\end{equation}
which yields
\begin{equation}
g_{\max} =
\frac{
\ln \left( n  N_{A} \right)
- \ln \left( N_{0} \right)
}{
\ln \left( N_{g} \right)
}.
\label{eq:eq13}
\end{equation}
In the special case where the cluster structure is preserved across generations, i.e., $N_{0} = N_{g}$, this expression simplifies to 
\begin{equation}
g_{\max} =
\frac{
\ln \left( n N_{A} \right)
}{
\ln \left( N_{g} \right)
}
- 1.
\label{eq:eq14}
\end{equation}

\section{Recurrence Relations Between Evolution Scales}

The connection between successive generations of clusters is established through recurrence relations for both the magnetization and the effective magnetic coupling. In this work, we adopt the following scale relation for the
magnetization:
\begin{equation}
F\!\left( m^{(g-1)} \right)
= \frac{m^{(g-1)}}{m_{\mathrm{sat}}},
\label{eq:eq15}
\end{equation}
where $m_{\mathrm{sat}} = 1.5$ corresponds to the saturation magnetization of a
spin-$3/2$ system. This choice ensures that the magnetic memory of the previous generation is preserved while keeping the magnetization properly normalized.

For the effective coupling constant, we introduce a recursive renormalization
relation of the form
\begin{equation}
J_{1}^{(g)} = a \, J_{1}^{(g-1)},
\label{eq:eq16}
\end{equation}
which leads to
\begin{equation}
J^{(g)} = a^{g} J^{(0)}.
\label{eq:eq17}
\end{equation}
The dimensionless coupling parameter is then given by
\begin{equation}
K^{(g)} = \frac{J^{(g)}}{k_{B} T}
= \gamma^{(0)} \frac{a^{g}}{T},
\label{eq:eq18}
\end{equation}
where the parameter $a$ is used to adjust the model to the physical system under investigation.

Accordingly, the parameter $a$ controls how the effective magnetic coupling evolves between generations. Furthermore, there is a relevant value of $a$ denoted as $a_{v}$ for the present work. The parameter $a_v$ is defined as the value of the scale factor $a$ at which the internal energy $u$ has a point of inflection (so $u'(a=a_{v})$ has a local extreme). There, there is a continuous phase transition where the specific heat reaches its maximum (for finite systems). However, in the thermodynamic limit the specific heat diverges (it corresponds to a kink in the slope of $u(T_{c})$). \\
Therefore, determining $a_{v}$ provides a way to identify a continuous phase transition. It is analogous to the fixed point in renormalization-group approaches.

\section{Robustness of the Recurrence Ansatz}

The proposed formalism relies on the introduction of a recurrence function $F(m)$ that relates the effective magnetization between successive cluster generations. In the simplest implementation, this function was chosen as
$F(m)=m/m_{\mathrm{sat}}$. In order to justify this choice and assess its robustness, we could generalize the recurrence ansatz to a one-parameter family $F_\alpha(m)$, which is given by 

\begin{equation}
F_\alpha(m) = \left( \frac{m}{m_{\mathrm{sat}}} \right)^\alpha ,
\label{eq:Falfa}
\end{equation}
which reduces to the original ansatz for $\alpha=1$.

By exploring numerically the exponent $\alpha$ under physically motivated constraints, it can be found that only a restricted range of $\alpha$ leads to thermodynamically consistent solutions that can be calibrated in agreement with experimental data (we found values in  $0< \alpha < 2$). This analysis justifies the selection of the linear case $\alpha=1$ as the natural baseline of the model and it emerges as the simplest choice.

It is important to highlight that the generalized recurrence function $F_\alpha(m)$ is required to satisfy a set of basic physical and structural conditions. First, the paramagnetic point must be preserved, implying $F_\alpha(0)=0$. Second, the function must saturate at the maximum magnetization, $F_\alpha(m_{\mathrm{sat}})=1$, preventing
unphysical growth of the order parameter. Third, $F_\alpha(m)$ must be a monotonically increasing function, ensuring that larger local magnetization leads to stronger effective ordering.

\section{Results}

Before discussing the numerical results, it is worth emphasizing the role of the key parameters of the multiscale cluster framework. The parameter $g_{max}$ controls the effective system size, while the scale factor $a$  encodes the renormalization of the magnetic coupling across generations. Together, these parameters determine how finite-size effects and critical behavior emerge in the model.

We consider clusters of size $N_{0}=N_{g}=4$ sites (see Fig.\ref{fig:fig1}) for the
ferromagnetic system defined on a lattice with honeycomb topology. Recently,
Gudelli \textit{et al.}~\cite{27} investigated the ferromagnetic material
CrI$_3$ for several multilayer configurations, including different types of
exchange interactions between magnetic neighbors. In the two-dimensional
monolayer (ML) case, the authors reported the nearest-neighbor magnetic
exchange parameter, which in our notation corresponds to $J^{(0)}$, of the
order of $6.91$~meV $= 1.107 \times 10^{-21}$~J. Therefore, we adopt the
reference value $\gamma^{(0)} \approx 10^{2}$~K as an input parameter in the
modeling.

\begin{figure}
    \centering
    \includegraphics[width=0.8\linewidth]{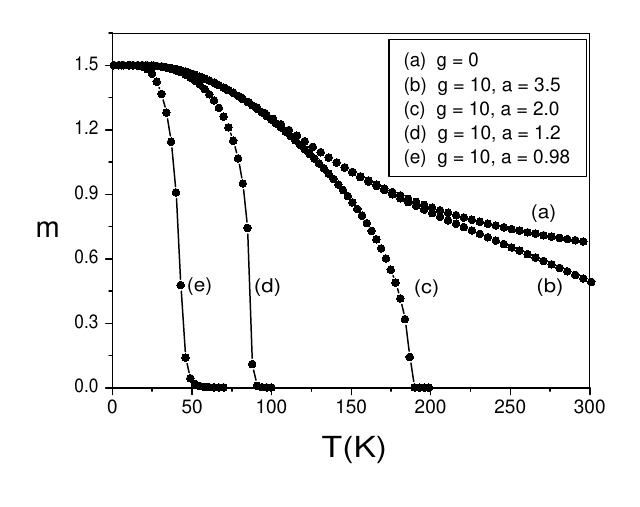}
    \caption{ Magnetization versus temperature for $g=0$ and for diferent values of the parameter $a$, where $g=10$.}
    \label{fig:fig2}
\end{figure}

Figure~\ref{fig:fig2} shows the magnetization as a function of temperature, $m(T)$, for the zeroth cluster generation ($g=0$) as well as for higher generations $g\geq 1$. In the curve labeled by (a), corresponding to $g=0$, the magnetization does not exhibit critical behavior, as expected for a theoretical treatment of a finite system
of size $N$. In contrast, curves from (b) to (e) present the case $g_{\max}=10$. For both $a>1$ and $a<1$, critical-like behavior of the magnetization is observed,
with $m \simeq 0$ above a characteristic temperature for the cases $a=2.0$,
$1.2$, and $0.98$. These results indicate that within the present modeling
framework it is possible to capture the criticality of the magnetic system,
a phenomenon that is observed in real materials.

Therefore, Figure~\ref{fig:fig2} illustrates the emergence of critical-like behavior induced by the
multiscale cluster-growth procedure. While the zeroth-generation cluster ($g=0$) corresponds to a finite system that does not exhibit magnetic
criticality, higher generations effectively represent increasingly larger 
systems. As the number of generations increases, long-range correlations are
progressively built into the model, allowing the magnetization to vanish
continuously at a characteristic temperature for appropriate values of the
scale parameter $a$. This behavior is reminiscent of real-space
renormalization-group flows, in which the effective coupling strength controls the stability of the ordered phase.

\begin{figure}
    \centering
    \includegraphics[width=0.9\linewidth]{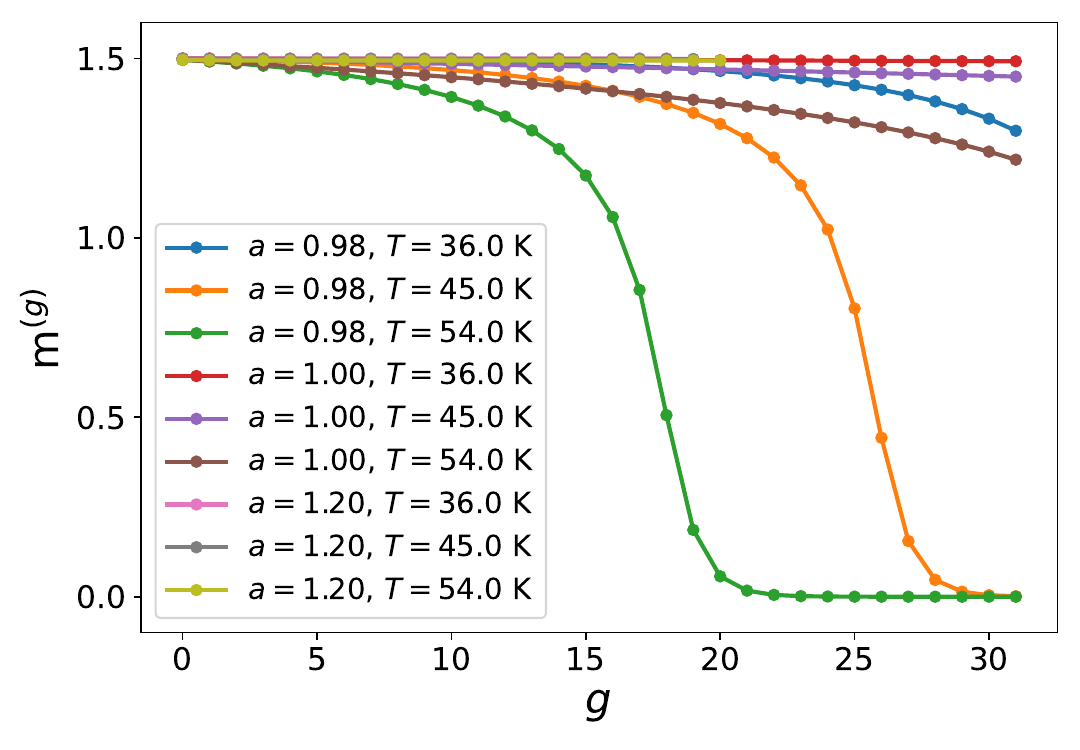}
    \caption{Scale evolution of the magnetization $m^{(g)}(T;a)$ as a function of the generation index $g$, obtained from the canonical partition sum of the cluster.
Results are shown for three representative temperatures ($T=0.8T_c$, $T=T_c$,and $T=1.2T_c$) and for different values of the scale parameter $a$. Note that here we have $g_{max}=31$.}
    \label{fig:fig2b}
\end{figure}

On the other hand, while Fig.\ref{fig:fig2} displays the temperature dependence of the magnetization for a
fixed generation, Fig.\ref{fig:fig2b} reveals the underlying scale evolution that leads to those kind of curves. The analysis explicitly shows three regimes controlled by the scale parameter $a$, in which there is a critical value of $a$ denoted by $a_{v}$. For $a < a_{v}$, the magnetization decays toward  $m^{(g)}\to 0$ as the generation index increases; (ii) for $a > a_{v}$, the magnetization grows and saturates, signaling a stable ordered phase; and for $ a \simeq a_{v}$, the slow decay of $m^{(g)}$ indicates a crossover between ordered and disordered magnetic regimes.

However, in this preliminary analysis we arbitrarily determined the values of  $g_{\max}$ and $a$. If the goal is to model a specific material, the parameters must be constrained by using experimental information for that material. This is dealt with in the following subsection. 

\subsection{Case study: the compound CrI$_3$}

From experimental data, the magnetic material CrI$_3$ has a mass density
$\rho \simeq 5.32$~g/cm$^{3}$ and a molar mass $m_{m} \simeq 432.71$~g/mol. Here we consider a hypothetical sample consisting of a single magnetic-domain region with a well-defined volume $V$.

For each volume scale $V$, one can estimate the corresponding mass $m$, number of mols $n$, number of molecules $N$, and the maximum number of generations $g_{\max}$ (see Eq.(\ref{eq:eq14})) required to represent that volume within the proposed formalism:

\noindent
(a) For a volume of order $V \equiv 1$~mm$^{3}$:
\begin{align}
m &=& \rho V = 5.32\times 10^{-3}\ \text{(g)}, \tag{22}\\
n &=& \frac{m}{m_{m}} = 1.23\times 10^{-5}\ \text{(mol)}, \tag{23}\\
N &\simeq& 7\times 10^{18}\ \text{(molecules)}, \tag{24}\\
g_{\max} &\simeq& 31\ \text{(generations)}. \tag{25}
\end{align}

\noindent
(b) For a volume of order $V \equiv 1\ \mu\text{m}^{3}$:
\begin{align}
m &=& \rho V = 5.32\times 10^{-12}\ \text{(g)}, \tag{26}\\
n &=& \frac{m}{m_{m}} = 1.23\times 10^{-14}\ \text{(mol)}, \tag{27}\\
N &\simeq& 7\times 10^{9}\ \text{(molecules)}, \tag{28}\\
g_{\max} &\simeq& 16\ \text{(generations)}. \tag{29}
\end{align}

\noindent
(c) For a volume of order $V \equiv 1$~nm$^{3}$:
\begin{align}
m &=& \rho V = 5.32\times 10^{-21}\ \text{(g)}, \tag{30}\\
n &=& \frac{m}{m_{m}} = 1.23\times 10^{-23}\ \text{(mol)}, \tag{31}\\
N &\simeq& 7\ \text{(molecules)}, \tag{32}\\
g_{\max} &\simeq& 1\ \text{(generations)}. \tag{33}
\end{align}

Note that for a volumetric sample with $V \equiv 1$~nm$^{3}$, the number of
particles is $N \simeq 7$, which is comparable to the size of the
zeroth-generation cluster ($N_{0}=4$).

Another important experimental quantity for monolayer CrI$_3$ is the phase
transition temperature (ferromagnetic/paramagnetic), which is approximately
$T_{f} \simeq 45$~K. This value can be obtained from the temperature at which
the constant-volume specific heat exhibits a maximum~\cite{27}. This
information is essential for determining the parameter $a$ within the present
formalism.

\subsection{Analysis of the phase transition}

One of the key experimental inputs for CrI$_3$ is its phase transition temperature, inferred from the maximum of the constant-volume specific heat, namely $T_{f} \simeq 45$~K~\cite{27}. We use this information to test the
modeling framework by fixing the temperature at $T=45$~K and studying the internal energy as a function of the scale parameter $a$, together with its derivative.

\begin{figure}
    \centering
    \includegraphics[width=0.9\linewidth]{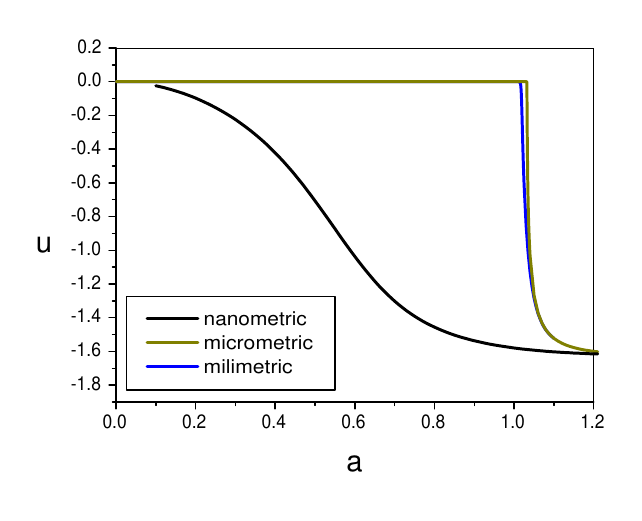}
    \caption{ Internal Energy  versus $a$ for
 the volumetric scales of the magnetic domains adopted in the modeling.}
    \label{fig:fig3}
\end{figure}

\begin{figure}
    \centering
    \includegraphics[width=0.9\linewidth]{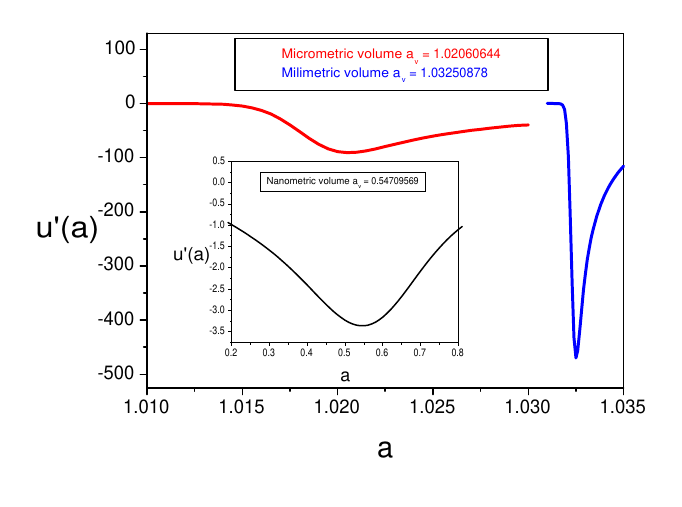}
    \caption{The behavior of the derivative $u\prime (a)$ is shown in order to determine the minima of these derivatives. The case for the nanometric volume is shown in the inset.}
    \label{fig:fig4}
\end{figure}

Figures~\ref{fig:fig3}--\ref{fig:fig4} show the results for $u(a)$ and its derivatives $u'(a)$ for the
three volume scales considered. In Fig.~\ref{fig:fig3}, the internal energy $u$ exhibits inflection points; from the derivatives shown in Fig.~\ref{fig:fig4} we determine the
 points $a_{v}$ where $u\prime$ exhibits its minima.

\begin{figure}
    \centering
    \includegraphics[width=0.9\linewidth]{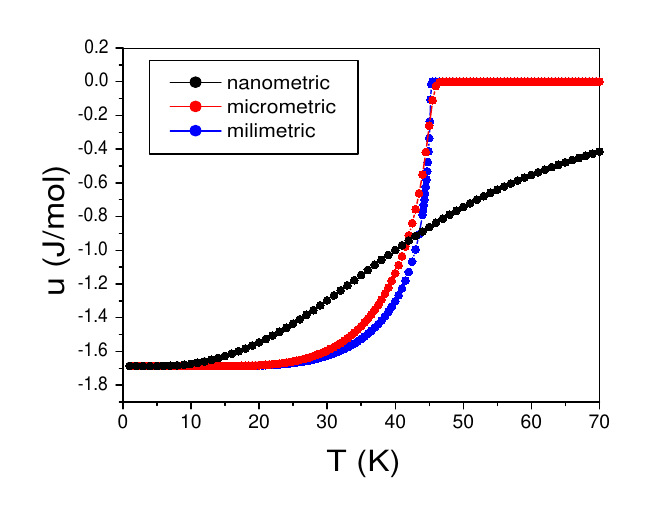}
    \caption{ Internal Energy versus temperature for
 the volumetric scales of the magnetic domains adopted in the modeling ( $V=1{nm}^{3}, V=1\ {\mu m}^{3},V=1{mm}^{3}$). }
    \label{fig:fig5}
\end{figure}

\begin{figure}
    \centering
    \includegraphics[width=0.9\linewidth]{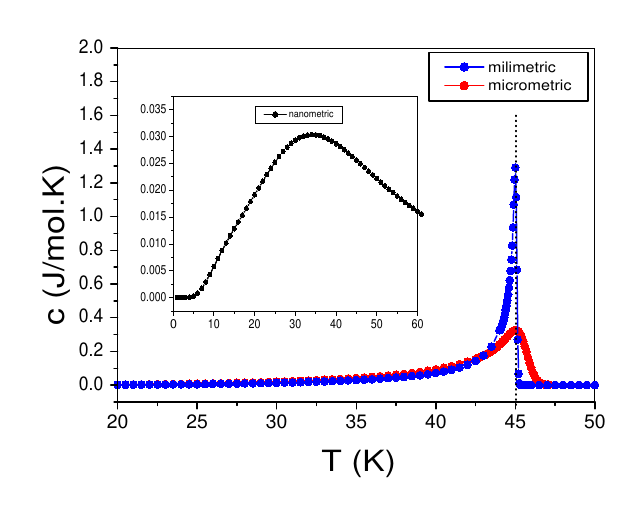}
    \caption{ Behavior of specific heat versus temperature for the adopted volume cases, obtained from the fixed values for $a = a_{v}$. The inset shows the nanometric volume case. }
    \label{fig:fig6}
\end{figure}

Subsequently, we solve the model by fixing $a=a_{v}$ for each volume $V$ and compute the internal energy and the specific heat as functions of temperature. The corresponding results are shown in Figs.~\ref{fig:fig5}-\ref{fig:fig6}, respectively.

Fig. \ref{fig:fig5} displays the characteristic behavior of the internal energy for each
magnetic-domain volume $V$. At low temperatures the energy approaches a
constant value, $u(T=0) = u_{0}$ (J\,mol$^{-1}$), while at high temperatures the system approaches $u=0$, which corresponds to the paramagnetic phase. Inflection points are also observed at $u(T_{c})$, which coincide with the temperatures at which the specific heat reaches its maximum.

On the other hand Figure~\ref{fig:fig6} shows the specific heat as a function of temperature. For $V=1\ \mu\text{m}^{3}$, a smooth maximum occurs at $T_{c}=45$~K, whereas for
$V=1$~mm$^{3}$ the maximum becomes sharper. In contrast, for $V=1$~nm$^{3}$ one finds $T_{c}\neq 45$~K.

\begin{figure}
    \centering
    \includegraphics[width=0.9\linewidth]{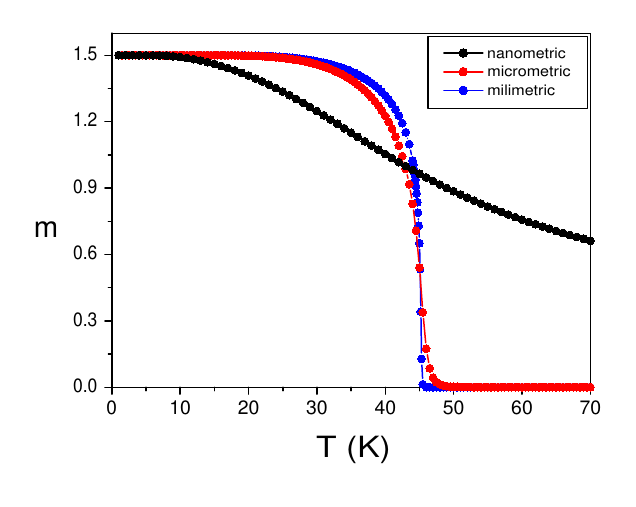}
    \caption{ Behavior of magnetization versus temperature for the adopted volume cases ( $V=1{nm}^{3}, V=1\ {\mu m}^{3},V=1{mm}^{3}$), obtained from the fixed values for $a = a_{v}$. }
    \label{fig:fig7}
\end{figure}

We therefore conclude that the working hypothesis that the specific-heat maximum can be captured by setting $a=a_{v}$ (obtained from the minimum of $u'(a)$) holds only when the magnetic system is composed of a sufficiently large number of particles. This is expected since for small $N$, as in the
case $V=1$~nm$^{3}$, the system does not display critical behavior in the magnetic order parameter ($m\simeq 0$) near $T_{c}$.
The magnetization curves $m(T)$ are shown in Fig.~\ref{fig:fig7} for each volume $V$. For $V=1$~nm$^{3}$, the magnetization does not approach a critical value ($m\simeq 0$) near $T=T_{c}$. In contrast, for $V=1\ \mu\text{m}^{3}$ and $V=1$~mm$^{3}$ the magnetization exhibits critical behavior, approaching 
$m\simeq 0$ near $T=T_{c}$. These results indicate, as expected, that ferromagnetic systems with small $N$ do not display criticality in $m$, whereas for very large $N$ the onset of critical behavior becomes apparent.

\begin{figure}
    \centering
    \includegraphics[width=0.9\linewidth]{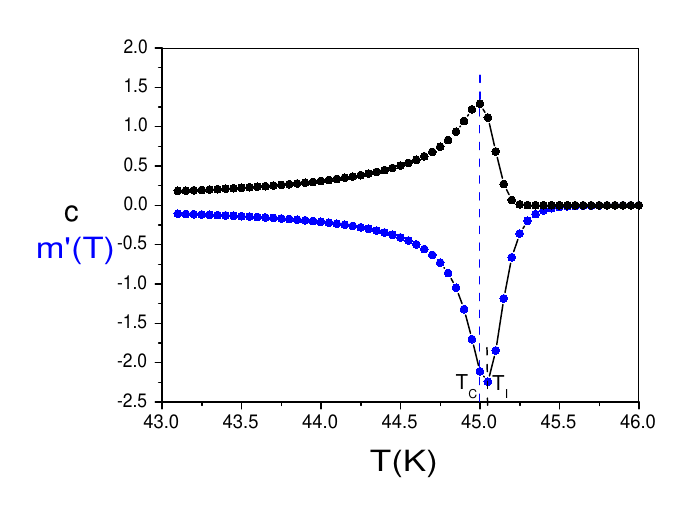}
    \caption{ Comparison of specific heat (curve with positive values) with the derivative of magnetization $m\prime (T)$ (curve with negative values) for the case of millimetric volume. }
    \label{fig:fig8}
\end{figure}

Since the magnetization curves exhibit inflection points, the first derivative
$m'(T)$ can be used to estimate the corresponding inflection temperature $T_{i}$. Fig. \ref{fig:fig8} shows results for the millimetric volume, where the inflection point is found at $T_{i}\simeq 45.048$~K, which is close to the
temperature of the specific-heat maximum, $T_{c}=45$~K. A similar behavior is observed experimentally for CrI$_3$ and is also found for the micrometric volume.

The qualitative agreement of $m(T)$ for $V=1\ \mu\text{m}^{3}$ and $V=1$~mm$^{3}$ with experimental results for CrI$_3$ is satisfactory~\cite{22}.
In particular, the magnetization approaches $m=0$ smoothly, as expected for a
system with very large but finite $N$. This supports that the present modeling
proposal is consistent with experimental observations.

\subsection{Entropy behavior}

Within the canonical ensemble formalism, the connection between statistical mechanics and thermodynamics is established through the Helmholtz free energy per particle, 
\begin{equation}
f = -\frac{1}{N_{c}} \frac{1}{K} \ln \left( Z_{N_{c}} \right),
\tag{34}
\end{equation}
where $f$ denotes the Helmholtz free energy density. Using a Legendre transformation, one obtains
\begin{equation}
f = u - Ts,
\tag{35}
\end{equation}
where $s$ is the entropy per particle. Therefore, the entropy can be computed
as
\begin{equation}
s = \frac{1}{T}\left( u - f \right),
\tag{36}
\end{equation}
or equivalently from the thermodynamic derivative
\begin{equation}
s = -\left( \frac{\partial f}{\partial T} \right).
\tag{37}
\end{equation}

\begin{figure}
    \centering
    \includegraphics[width=0.9\linewidth]{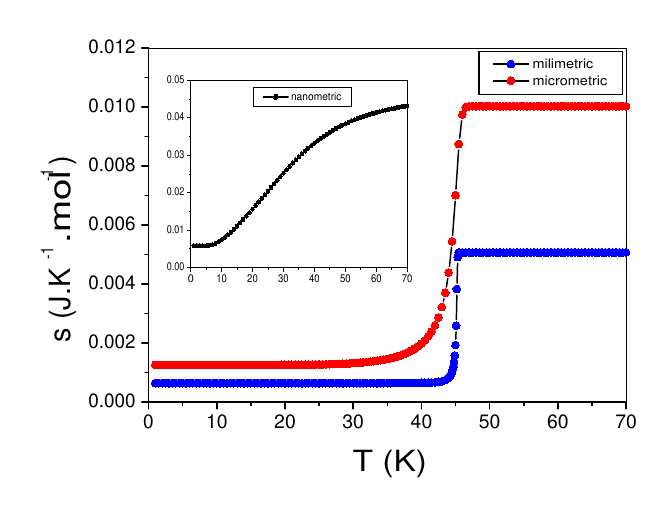}
    \caption{Behavior of entropy versus temperature for the cases of adopted magnetic domain volumes.  }
    \label{fig:fig9}
\end{figure}

Fig. \ref{fig:fig9} presents the entropy as a function of temperature for the three volumetric scales considered. At low temperatures, the entropy approaches a nonzero constant value, which is consistent with the third law of thermodynamics. Recall that Nernst's theorem states that the entropy of the
ground state vanishes if it is non-degenerate. In the present ferromagnetic system the ground state is twofold degenerate ($m=\pm 1.5$), implying a residual entropy at low temperature.

\subsection{Determination of critical exponents}

In the critical regime ($T\rightarrow T_{c}$), asymptotic theory predicts that
the magnetization behaves as
\begin{equation}
m(t) = A |t|^{\beta}, \qquad t \rightarrow 0,
\tag{38}
\end{equation}
or equivalently
\begin{equation}
\ln(m) = \ln(A) + \beta \ln\!\left(|t|\right),
\tag{39}
\end{equation}
where the reduced temperature is defined as
\begin{equation}
t = \frac{T_{c}-T}{T_{c}}.
\tag{40}
\end{equation}

\begin{figure}
    \centering
    \includegraphics[width=0.9\linewidth]{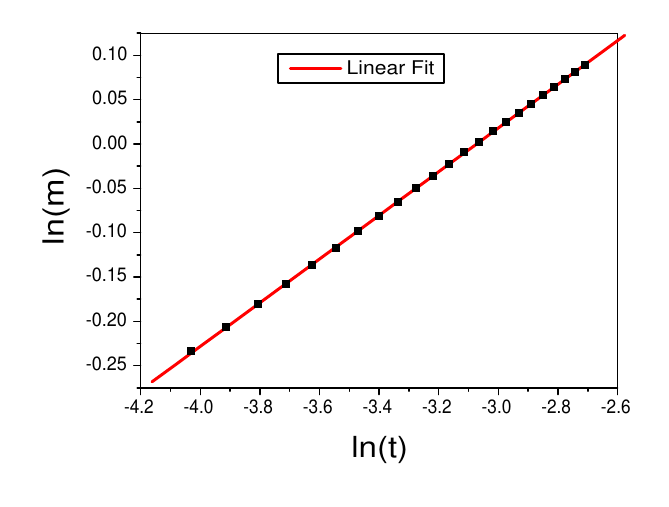}
    \caption{Linear fit adjustment obtained from the critical window based on the \( m(t) \) relation of the asymptotic regime for the micrometric volume, yielding the critical exponent \( \beta = (0.2465 \pm 3.6561) \times 10^{-4} \).  }
    \label{fig:fig10}
\end{figure}

\begin{figure}
    \centering
    \includegraphics[width=0.9\linewidth]{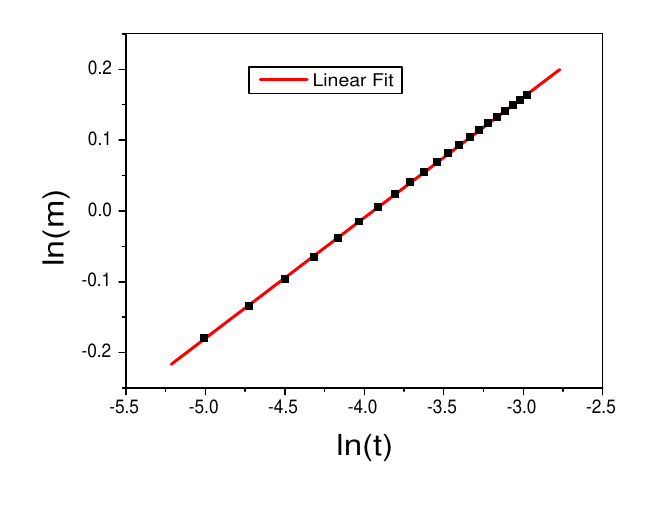}
    \caption{ Linear fit adjustment obtained from the critical window based on the \( m(t) \) relation of the asymptotic regime for the millimetric volume, yielding the critical exponent \( \beta = (0.1703 \pm 4.5955) \times 10^{-4} \). }
    \label{fig:fig11}
\end{figure}

In practice, one must identify, in the graphical analysis of $\ln(m)$ versus
$\ln(t)$, the most reliable critical window that satisfies both the regime
$t\rightarrow 0$ and the linear behavior implied by Eq.~(39). Figures~\ref{fig:fig10}-\ref{fig:fig11} show the corresponding plots for the cases where $V=1\ \mu\text{m}^{3}$ and
$V=1$~mm$^{3}$. From linear fits we obtain the following values for the critical exponent:
\begin{eqnarray*}
\beta = 0.2465(4),\qquad V=1\ \mu\text{m}^{3},\\
\beta = 0.1703(5), \qquad V=1\ \text{mm}^{3}.
\end{eqnarray*}

For $V=1$~mm$^{3}$, the optimal critical window was identified in the interval
$42.7 < T < 44.7$, where $T=44.7$ was used as the reference point of the fit,
implying $t=0.006667$. For $V=1\ \mu\text{m}^{3}$, the window was identified in
the interval $42.0 < T < 44.2$, where $T=44.2$ implies $t=0.017778$.

From the values obtained for the critical exponent $\beta$, we conclude that
the present modeling is not consistent with a simple mean-field theory (MFT),
for which $\beta=0.5$. Instead, the results suggest a tendency toward an
Effective Field Theory (EFT) or Renormalization Group (RG) type behavior, as
initially proposed, where $\beta=0.125$ for the two-dimensional Ising
universality class. This tendency becomes more evident as the cluster volume
approaches the millimetric scale.

A Ph.D. thesis~\cite{28} suggests, based on experimental data and Monte Carlo simulations for monolayer CrI$_3$, that the experimental value of $\beta$ may lie in the range $0.13$--$0.15$, close to the expected two-dimensional Ising
value $\beta = 0.125$.


\section{Conclusions}
In this work, we propose a methodology for solving spin models, based on a process of gradual growth of the population of particles that compose the first-neighbor spin-3/2 Ising ferromagnetic system, located on a honeycomb lattice.

One of the merits of the proposal in this work is related to the simplification of the thermodynamic-statistical treatment of the studied physical system. For treating the modeling of a system of magnetic spins, in the case of spin 3/2, it would be necessary to apply $4^{N}$ magnetic states in the partition function within the formalism of the Canonical Ensemble of Statistical Mechanics. This calculation is an improbable task to perform for very large $N$, even using computational resources. However, with this proposal of dynamics for the evolution of the size ($N$) of the physical system from the construction of finite clusters ($N_{g}$), the calculation of the partition function is performed from $4^{N_{g}}$ magnetic states in each generation $g$. This makes the computational process much faster and the model can be solved, where the system size can be very large and given by $N=N_{0}\left( N_{g}\right) ^{g}$ particles. As can be observed, the system size grows in a progressive geometric manner with each generation $g$.

In this modeling proposal, we neglect the true quantity of accessible states ($4^{N}$) in the formalism. This modeling works adequately because we apply the resulting local magnetization for each generation $g$ as the variable of the magnetic site in the Hamiltonian that represents the system of the next generation $g+1$ of the magnetic cluster. This modeling has an aspect of Effective Field Theory (EFT), as it is more concerned with the magnetic interaction between the sites that compose the physical system.

One of the objectives of this work was to study the thermodynamics and magnetization of the real material CrI${3}$ (Chromium Triiodide), which is one of the first real 2D Ising-type magnetic systems observed experimentally. The results obtained for magnetization $m(T)$ present similarities to the results observed experimentally. Among the results, we have the presence of an inflection point for magnetization, which is close to the peak of the specific heat $c{v}(T)$, indicating the phase transition point of the ferromagnetic system. The criticality is smoothed in the approximation of $m\simeq 0$ and at the maximum of the specific heat. These behaviors are typical of finite systems, as is the case with real materials.

We apply the Canonical Ensemble formalism in the modeling, starting from an exact case for a finite cluster with $N_{0}=4$ sites in the zero cluster generation ($g=0$). Subsequently, for generations $g\geq 1$, we apply the Canonical Ensemble by substituting the spin variables in the theoretical model with the representation that each site is the resultant of a cluster with $N_{g-1}$ sites from the previous generation $g-1$. The cluster generations $g$ present clusters in the form of a hierarchical lattice. We also have that the number of generations ($g$) was determined from the scale of the size of the adopted magnetic domain volume. The results demonstrate that even for a simple cluster with $N_{0}$ and $N_{g}$ being small, very good results can be obtained in qualitative and quantitative aspects when compared to results of real materials.

The model was calibrated using the experimental data of the phase transition temperature for the CrI$_{3}$ compound, in the monolayer structure, which presents the coupling factor $J^{(0)} = 1.107\times 10^{-21}$ J and critical temperature $T\simeq 45$ K, verified from the point of maximum in the specific heat behavior of this material.

From this model calibration, we can determine the theoretical curve of specific heat $c(T)$, obtaining the phase transition temperature ($T_{c}$) at the maximum of this thermodynamic property. This result indicates that the form of the model calibration is correct, as it allowed the modeling to obtain the same result for the phase transition point of the ferromagnetic system studied theoretically.

We also studied the disorder of the system from the entropy $s(T)$ of the magnetic system. At low temperature, a residual entropy ($s\neq 0$) was obtained; this occurs due to the system being doubly degenerate ($m=\pm 1.5$) in this temperature regime. This behavior is in accordance with the Third Law of Thermodynamics.

We expect to apply this present theoretical resolution to magnetic systems that present several coupled magnetic domains, exhibiting symmetries of the Ising and Heisenberg type composed of various anisotropies. This will allow for the theoretical study of real magnetic systems with more complex structures.

\section*{Acknowledgments}
The authors acknowledge the financial support from the National Council for Scientific and Technological Development (CNPq) - Brazil.





\section*{References}

\begin{thebibliography}{99}
\bibitem{1} L. Onsager, Phys. Rev. 65, 117 (1944).

\bibitem{2} N. Goldenfeld, Lectures on Phase Transitions and the
Renormalization Group (Addison-Wesley, Reading, MA, 1992).

\bibitem{3} K. G. Wilson and J. Kogut, Phys. Rep. 12, 75 (1974).

\bibitem{4} R. Kikuchi, Phys. Rev. 81, 988 (1951).

\bibitem{5} D. C. Mattis, The Theory of Magnetism (Springer, Berlin, 1988).

\bibitem{6} N. W. Ashcroft and N. D. Mermin, Solid State Physics (Saunders
College, Philadelphia, 1976).

\bibitem{7} D. P. Landau and K. Binder, A Guide to Monte Carlo Simulations in Statistical Physics (Cambridge University Press, Cambridge, 2000).

\bibitem{8} U. Wolff, Phys. Rev. Lett. 62, 361 (1989).

\bibitem{9} R. H. Swendsen and J.-S. Wang, Phys. Rev. Lett. 58, 86 (1987).

\bibitem{10} A. W. Sandvik, Computational Studies of Quantum Spin Systems, AIP Conf. Proc. 1297, 135 (2010).

\bibitem{11} U. Schollw\"{o}ck, Rev. Mod. Phys. 77, 259 (2005).

\bibitem{12} P. Weiss, J. Phys. Theor. Appl. 6, 661 (1907).

\bibitem{13} C. G. Shull et al., Phys. Rev. 76, 1256 (1949).

\bibitem{14} R. L. M\"{o}ssbauer, Z. Phys. 151, 124 (1958).

\bibitem{15} J. Clarke and A. I. Braginski, The SQUID Handbook (Wiley-VCH, Weinheim, 2004).

\bibitem{16} J. W. Lynn, J. Appl. Phys. 57, 8 (1985).

\bibitem{17} R. von Helmolt et al., Phys. Rev. Lett. 71, 2331 (1993).

\bibitem{18} A. Hubert and R. Sch\"{a}fer, Magnetic Domains (Springer, Berlin, 1998).

\bibitem{19} Y. Martin and H. K. Wickramasinghe, Appl. Phys. Lett. 50, 1455
(1987).

\bibitem{20} K. A. McEwen, J. Magn. Magn. Mater. 129, 94 (1994).

\bibitem{21} R. Coldea et al., Science 327, 177 (2010).

\bibitem{22} B. Huang et al., Nature 546, 270 (2017).

\bibitem{23} C. Gong et al., Nature 546, 265 (2017).

\bibitem{24} Y. Zhang et al., Phys. Rev. Lett. 123, 047202 (2020).

\bibitem{25} J. R. Viana et al., J. Magn. Magn. Mater. 369, 101 (2014).

\bibitem{26} J. R. Viana et al., Int. J. Mod. Phys. B 31, 1850038 (2017).

\bibitem{27} V. K. Gudelli and G.-Y. Guo, New J. Phys. 21, 053012 (2019).

\bibitem{28} M. Augustin, Multiscale Approaches on Two-Dimensional Magnetic Materials: From Atomistic to Mesoscale Level, Ph.D. thesis, Queen's University Belfast (2021).
\end{thebibliography}
\end{document}